\documentclass[prl
,twocolumn%
,superscriptaddress%
,showpacs]{revtex4}
\usepackage{graphicx}%
\usepackage{dcolumn}
\usepackage{amsmath}

\makeatletter
\def\btt#1{\texttt{\@backslashchar#1}}%
\DeclareRobustCommand\bblash{\btt{\@backslashchar}}%
\makeatother

\begin{document}
\bibliographystyle{apsrev}
\title{\textit{Ab initio} studies of the spin-transfer torque in
  tunnel junctions} 
\author{Christian Heiliger}%
 \email{christian.heiliger@nist.gov}
\affiliation{%
Center for Nanoscale Science and Technology, National Institute of
 Standards and Technology, Gaithersburg, MD 20899-6202 
}
\affiliation{%
Maryland NanoCenter, University of Maryland, College Park, MD, 20742
}
\author{M. D. Stiles}%
\affiliation{%
Center for Nanoscale Science and Technology, National Institute of
 Standards and Technology, Gaithersburg, MD 20899-6202 
}
%
\date{\today}
\begin{abstract}
We calculate the spin-transfer torque in Fe/MgO/Fe tunnel junctions
and compare the results to those for all-metallic 
junctions. We show that the spin-transfer
torque is interfacial in the ferromagnetic layer to a greater degree 
than in all-metallic 
junctions. This result originates in the half metallic
behavior of Fe for the $\Delta_1$ states at the Brillouin zone
center; in contrast to all-metallic structures, dephasing does not play
an important role.  We further show that it is possible to get a
component of the torque that is out of the plane of the magnetizations
and that is linear in the bias.  However, observation of such a
torque requires highly ideal samples.  In samples with typical
interfacial roughness, the torque is similar to that in all-metallic
multilayers, although for different reasons.
\end{abstract}
%
\maketitle
%
%
%

The discovery of giant magnetoresistance (GMR) in
spin-valve systems~\cite{binasch89,baibich88} and the rediscovery of
the tunneling magnetoresistance (TMR) in tunnel
junctions~\cite{moodera95,miyazaki95} has led to 
applications such as hard-disk read heads, sensors, and storage
elements in magnetic random access memory (MRAM).  Both GMR and TMR
occur in junctions in which two ferromagnetic leads are separated by a
spacer layer, which is a non-magnetic metal in case of a GMR junction
and a tunnel barrier for a TMR junction. The resistance $R$ and
therefore the conductance $g$ of these junctions are a function of the
relative angle $\theta$ between the magnetizations of the
ferromagnetic leads. The magnetoresistance (MR) ratio is 
$[R(180^\circ)-R(0^\circ)]/R(0^\circ)$.  Typical GMR ratios are in the
range of $50\ \%$ \cite{heiliger06_mat} and can be explained by spin
dependent interface scattering in a semiclassical
picture~\cite{levy94,gijs97,zahn98}.  TMR ratios
can exceed several hundred percent in crystalline Fe/MgO/Fe tunnel
junctions~\cite{yuasa04a,parkin04} as predicted
theoretically~\cite{butler01,mathon01}. 

These high TMR values in crystalline Fe/MgO/Fe junctions originate
in the symmetry-dependent transmission probabilities through the MgO barrier
at the Brillouin zone center ($\overline{\Gamma}$ point) combined with
the exchange splitting of the Fe band
structure~\cite{butler01,heiliger06_mat}. States which have the full
rotational symmetry of the interface, said to have $\Delta_1$
symmetry, decay the most slowly in MgO and hence dominate the
tunneling current.  At the Fermi level, Fe is a half metal at the
$\overline{\Gamma}$ point for states with $\Delta_1$ symmetry, having
only majority states. The half metallic nature of
the states that dominate the tunneling leads to a much higher current
for parallel than for 
antiparallel alignment of the magnetizations.

Spin-transfer torque, an effect predicted by
Slonczewski~\cite{slonczewski89,slonczewski96} and
Berger~\cite{berger96}, can be used to switch the magnetic
orientation of ferromagnetic layers in GMR and TMR devices.  For
this purpose a current is driven through the sample and becomes spin
polarized in one ferromagnetic layer. This polarization persists going
through the spacer layer and entering the other ferromagnetic
layer. If the spins of the polarized current are not aligned
with the magnetization, they precess around it. This precession in
turn creates a torque on the 
magnetization and  can reverse
the magnetization if the current is high enough.  There is currently
significant interest in  understanding spin-transfer torques in
tunnel junctions as a way to allow the development of MRAM
applications~\cite{diao07}.

Spin transfer torques are well understood in all-metallic trilayer
structures~\cite{stiles06}.  There, the current is
carried by electrons over the whole Fermi surface.  In ferromagnetic layers
and at their interfaces, electrons precess at different
rates and the components of the spins transverse to the magnetization
rapidly become out of phase from each other.  Two properties of the
torque follow from the strong dephasing of the
electron spins~\cite{stiles02}. First, the spin transfer torque largely
occurs at the interfaces between ferromagnetic layers and non-magnetic
layers. Second, it is largely in the plane defined by the
magnetizations of the two ferromagnetic layers. These properties are used
in almost all modelling of dynamics of GMR devices~\cite{stiles06}
and have been used without change in the modelling of TMR devices.

\begin{figure}
\includegraphics[width=0.85 \linewidth]{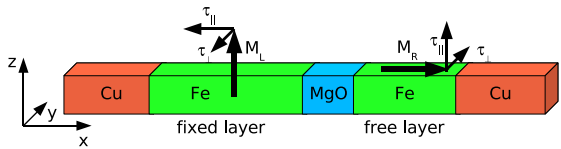}
\caption{ (Color online) Top: Schematic geometry. The left $\vec{M}_L$ and right
$\vec{M}_R$ magnetizations of the tunnel junction are taken to lie in in the $xz$
plane, at a relative angle $\theta$ (here taken to be $90\ ^\circ$).
The spin transfer torque acts perpendicular to each magnetization and
can be divided into the in-plane torque $\tau_\parallel$, which lies
in the $xz$ plane, and the out-of-plane torque $\tau_\perp$, which
points into the $y$ direction perpendicular to the plane defined by
$\vec{M}_L$ and $\vec{M}_R$.  }
\label{fig_sketch}
\end{figure}

In typical tunnel junctions, the current and spin
current are carried by a small fraction of the Fermi surface and
dephasing is greatly reduced.  Here, we use \textit{ab initio}
calculations to compute the spin 
transfer torques in Fe/MgO/Fe tunnel junctions.  We show that in spite
of the reduced dephasing, the torque is still approximately confined
to the interface and the in-plane torque is much larger than the
out-of-plane torque (see Fig.~\ref{fig_sketch}).

We treat the same structure as in previous
studies on Fe/MgO/Fe~\cite{heiliger07b} taking into account the
experimentally observed relaxation of the Fe layer next to the
interface~\cite{meyerheim01}. The junctions consist of an
Fe fixed layer, an MgO barrier,
and an Fe free layer embedded
between semi-infinite Cu in a bcc-Fe structure. The potentials are
calculated self-consistently using a screened
Korringa Kohn Rostoker multiple scattering Green's function
approach.
We obtain potentials for junctions with different free layer
thicknesses using the frozen potential
approximation and the potentials from a calculation with
20 Fe monolayers in each magnetic layer. 
We calculate the linear-response torque using the
magnetic moment $\delta\vec{m}(E_F)$ of those electrons that contribute to
the current. The torkance  $d\vec{\tau}/dV$ is the variation of the
torque $\vec{\tau}$ with the 
voltage $V$.  The torkance acting on layer $i$
is~\cite{haney07}
\begin{equation} \label{eq_torque}
  \frac{d \vec{\tau}_i}{dV} = \frac{\mu_B g_0}{2 e A} \vec{\Delta}_i  \times 
  \delta\vec{m}(E_F) \ , 
\end{equation}
where $\vec{\Delta}$ is the exchange field pointing in the direction
of the magnetization of the corresponding layer, $A$ is the area of
the in-plane unit cell, and $g_0={e^2}/{h}$ is the quantum of
conductance. For a 
description of our implementation of the non-equilibrium Green's
function technique see
Refs.~\onlinecite{heiliger07_arx,henk06,haney07}.  
The total non-equilibrium magnetization 
$\delta\vec{m}_i(E_F)=(1/2)[\delta\vec{m}^{\rm
    L}_i(E_F)-\delta\vec{m}^{\rm R}_i(E_F)]$ contains separate
contributions from  electrons incident from the left and holes incident from
right (for a positive bias).  

\begin{figure}
\includegraphics[width=0.85 \linewidth]{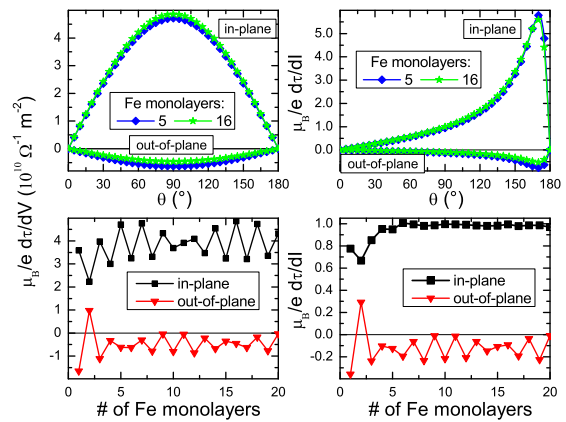}
\caption{
(Color online) Top: Torkance (left) and torque per current (right) acting on the free
layer as a function of the relative angle $\theta$ between the
magnetizations of the ferromagnetic electrodes. Results are shown
for different thicknesses of the Fe free layer as indicated in the figure.
Bottom: Torkance (left) and torque per current (right) as a function
of the free layer thickness for an angle of $90\ ^\circ$.  All
calculations have a 20 monolayer fixed layer and a 6 monolayer barrier.
}
\label{fig_torkance}
\end{figure}

Fig.~\ref{fig_torkance} shows our \textit{ab initio} calculations
of the torkance
as a function of the relative angle $\theta$
between $\vec{M}_L$ and $\vec{M}_R$ for different thicknesses of the
free layer.  Both the component in the plane of the two
magnetizations, $d\tau_\parallel/dV$, and the component out of the
plane $d\tau_\perp/dV$ are almost perfectly sinusoidal.  For current
biased applications, the torque per current 
$d\vec{\tau}/dI$ is of greater interest.  This is related to the torkance by 
the conductivity $g=dI/dV$, $g(d\vec{\tau}/dI)=d\vec{\tau}/dV$. In tunnel
junctions with very high TMR ratios the conductivity depends strongly
on the angle between the magnetizations so that $d\vec{\tau}/dI$ is
highly asymmetric.   The critical
voltages and currents for switching out of the parallel and
antiparallel states are proportional to the inverses of the slopes of
these curves at 
$\theta=0^\circ$ and $180^\circ$ respectively. 

The bottom left panel of Fig.~\ref{fig_torkance} shows that both the
in-plane and 
out-of-plane torkance oscillate as a function of the Fe free layer
thickness.  
The oscillations fit a sine curve with a period that is
incommensurate with the 
lattice spacing (hence the apparent beating of the amplitude). 
This period is very close to the Fermi wave vector of the $\Delta_1$
band in Fe at the $\overline{\Gamma}$ point.  
This agreement of the periods
indicates that the important states are located close to the
Brillouin zone center. The conductance shows similar
oscillations~\cite{heiliger07b}, so that the torque per current in the
right bottom panel of Fig.~\ref{fig_torkance} is largely independent
of the thickness.  However, there is a phase shift between
the oscillations in the in-plane and out-of-plane torkance so that the
oscillations in $d\tau_\perp/dI$ are even stronger than they are in
the torkance. 

Similar oscillations of the conductance and
magnetoresistance have been observed as a function of the thickness of
an additional nonmagnetic layer inserted next to the
barrier~\cite{yuasa02,itoh03}.  For Fe/MgO/Fe, the origin of these
oscillations is 
subtle.  As we show below, the torque is largely restricted to the
interface because the minority component is evanescent.  However, the
torque changes as a function of the free layer thickness due to
coherent effects: specifically, phase shifts in the majority channel due to
quantum well states.  

Note that in Fig.~\ref{fig_torkance} the torque per current is very
close to $\mu_{\rm B}/e$ when the magnetizations are at an angle of
$90\ ^\circ$.  This is the expected behavior for a simple model of a
junction between two half metals, for which the polarization of the
current is 100~\%.  Each electron spin that traverses the
junction rotates by $90^\circ$.  This change in angular momentum is
shared by the fixed and free layers.  Since the torque must be
perpendicular to the magnetizations, the change in angular momentum
supplied to each layer must be $\mu_{\rm B}$.
For greater angles, the torque per current can be
several factors higher, a result that seems counterintuitive.  In
fact, as the barrier is made thicker, there is no intrinsic limit to
how large the quantity can get.  This
behavior results from the contribution of electrons that penetrate the
barrier but reflect.  They contribute to the torkance but not to the
conductivity.  Consider a free layer with a magnetization at a
relative orientation of either $45^\circ$ or $135^\circ$ with respect
to that of the fixed layer.  In both cases, the
majority electrons that tunnel from the fixed layer have the same
transverse component and will exert roughly the same torque.  The
electrons tunneling into the free layer with a magnetization at
$45^\circ$ are approximately 85~\% majority and 15~\% minority in the
free layer while the electrons tunneling into the free layer with a
magnetization at $135^\circ$ are approximately 15~\% majority and
85~\% minority in the free layer.  The tunneling current will be much
greater in the former case than in the latter so that the ratio of
the torque to the tunneling current will be greater when the layers
are at $135^\circ$.

\begin{figure}
\includegraphics[width=0.85 \linewidth]{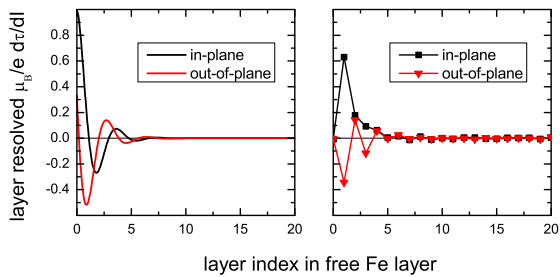}
\caption{ 
(Color online) Layer resolved torque per current acting on the free layer
at $\theta=90\ ^\circ$. Left: Free electron model using a simple
rectangular barrier (B) with an imaginary wavevector of
$k^\uparrow_B=k^\downarrow_B=4\ i\ \text{nm}^{-1}$ for majority
($\uparrow$) and minority spins ($\downarrow$) within the barrier.
Left ($L$) and right ($R$) of the barrier we use the half-metallic case
present in Fe at the $\overline{\Gamma}$ point. That is, the majority
wavevector is real $k^\uparrow_L=k^\uparrow_R=12\ \text{nm}^{-1}$,
whereas the minority wavevector is imaginary
$k^\downarrow_L=k^\downarrow_R=5\ i\ \text{nm}^{-1}$.  The thickness
of the barrier is $80\ \text{nm}$. Right: For comparison the
\textit{ab initio} result is shown integrated over the whole Brillouin
zone, for a free layer with 20 monolayers.  
}
\label{fig_layer}
\end{figure}

The weak influence of free layer thickness on $d\tau_\parallel/dI$ in
Fe/MgO/Fe and the 
saturation for more than three monolayers of Fe implies that the
torque is restricted to the layers next to the interface even though
there is only minimal dephasing. 
To understand this behavior, the right panel of Fig.~\ref{fig_layer}
shows our \textit{ab initio} 
results for the layer resolved torque.  
We analyze this result by comparison to a simple free
electron model in the left panel of Fig.~\ref{fig_layer}.

In the simple model, we use the wave vectors of the
$\Delta_1$ band in Fe at the Fermi level to model the half metallic
nature at the $\overline{\Gamma}$ point. 
In a half metal, the components of the non-equilibrium magnetization
perpendicular to the magnetization and hence the torque arise from coherent
interference between the majority and minority spin components.  Since
the minority state is evanescent, the torque must decay. 
After a
short distance only the component of the spin current
pointing along $\vec{M}_R$ is left.
The decay of the torque, which results from the half-metallic
character of the ferromagnetic electrodes, is even faster than in
spin-valve 
systems where the decay follows a power law resulting from
dephasing~\cite{stiles02}.
The
agreement between the {\it ab initio} results and the simple model
emphasizes that the states around the 
Brillouin zone center dominate the torque within the free
layer.  

During this decay, the electrons precess around $\vec{M}_R$ within the
$zy$ plane leading to decaying oscillations of the components of the
magnetic moments of the conduction electrons along the $z$ and $y$
directions.  These 
oscillations in
the magnetic moments are phase-shifted by $90\ ^\circ$ with respect to
each other \cite{kalitsov06}. They lead to
the oscillations in the torque 
components having the corresponding phase
shift as discussed in Fig.~\ref{fig_torkance}.  
Within a given layer, the sizes of
$\tau_\parallel$ and $\tau_\perp$ are comparable on average.  The
total torques are 
determined by the starting phase and the strength of the decay. 
In addition, Fig.~\ref{fig_layer} only shows the torque of
right-going electrons. 
The contributions of left-going holes have to be
added in order to obtain the torque in
Eq.~\ref{eq_torque}. 

There are additional contributions to the torque that are important for
thinner barriers 
but are negligible for 6 monolayer and thicker
barriers. 
In Fig.~\ref{fig_torkance}, there are oscillations that are barely visible,
which originate in 
states away from the Brillouin zone center for which Fe is not half-metallic.  
These
oscillations are long ranged due to reduced dephasing and could be
important for barriers that are thinner than those typically fabricated.  

Our calculations of the
bias dependence of the out-of-plane component of the torkance address
a topic of recent experimental interest.
Spin-transfer-driven ferromagnetic resonance (ST-FMR) quantitatively
measures the magnitude and direction of the spin transfer torque in
tunnel junctions~\cite{tulapurkar05,sankey06,sun07}.  Tulapurkar
\textit{et al.}~\cite{tulapurkar05} measured a linear dependence of
$\tau_\perp$ on the applied bias for small voltages (non-zero
$d\tau_\perp/dV$) whereas Sankey \textit{et al.}~\cite{sankey07}
measured $\tau_\perp$ to be linearly independent of $V$.

This latter result is consistent with theoretical 
arguments~\cite{slonczewski05,theodonis06} that hold in special cases.
For junctions that are symmetric about the center of the barrier and 
in which electrons leaving the ferromagnetic layers into the leads 
are aligned with the magnetizations, the left going electrons cause 
the same out-of-plane torque in the free layer as the right going 
ones, that is  $\tau_\perp^L(E_F)=\tau_\perp^R(E_F)$. 
When the outgoing electrons are aligned with the magnetizations, the
total torque on both layers in a symmetric junction must be in-plane.
If this condition holds, Eq.~\ref{eq_torque} gives
$d\tau_\perp/dV=0$  and there is no linear
dependence on the applied voltage for small biases. 
However, for asymmetric junctions these arguments do not apply.  

\begin{figure}
\includegraphics[width=0.4 \linewidth]{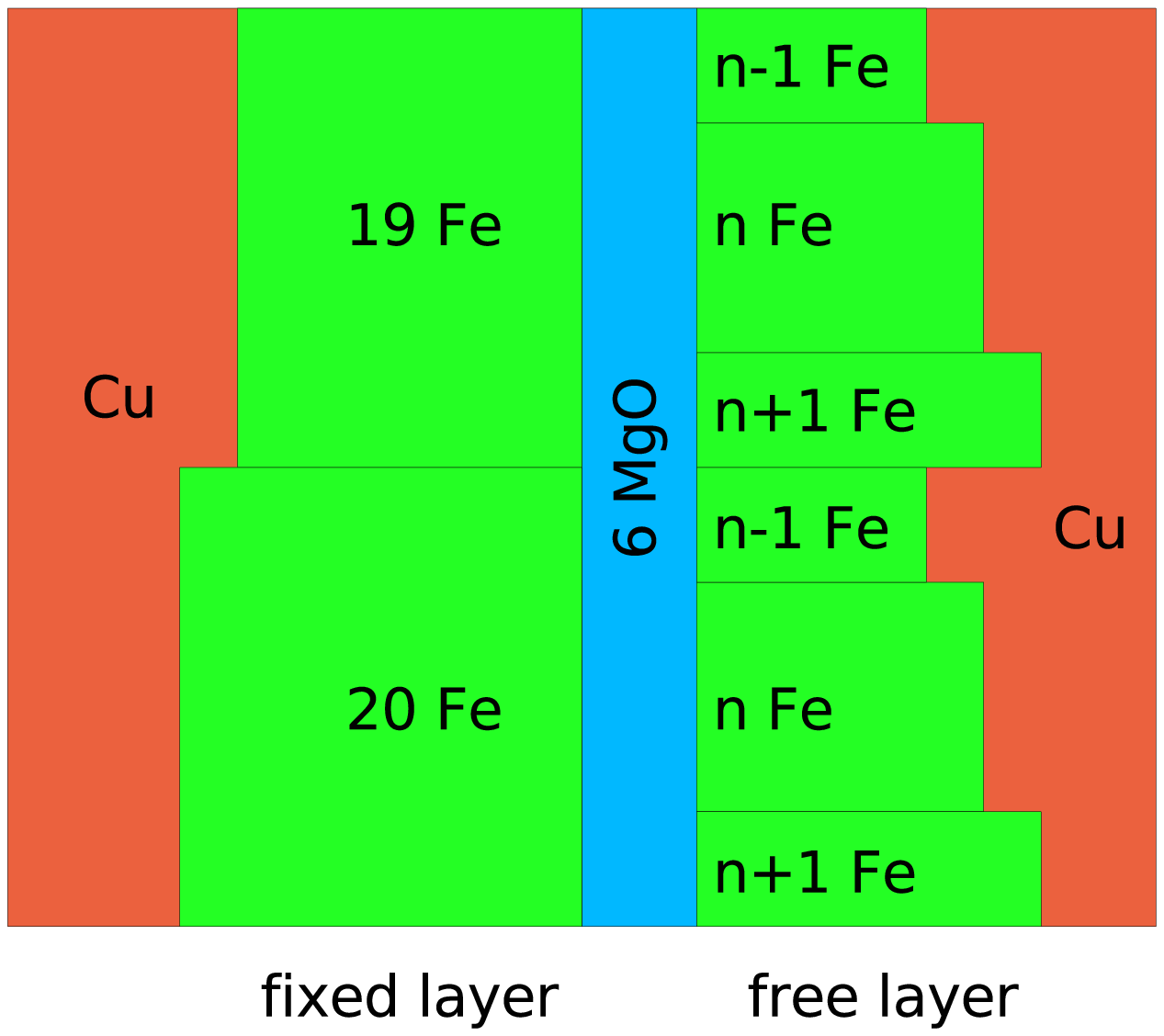}
\includegraphics[width=0.4 \linewidth]{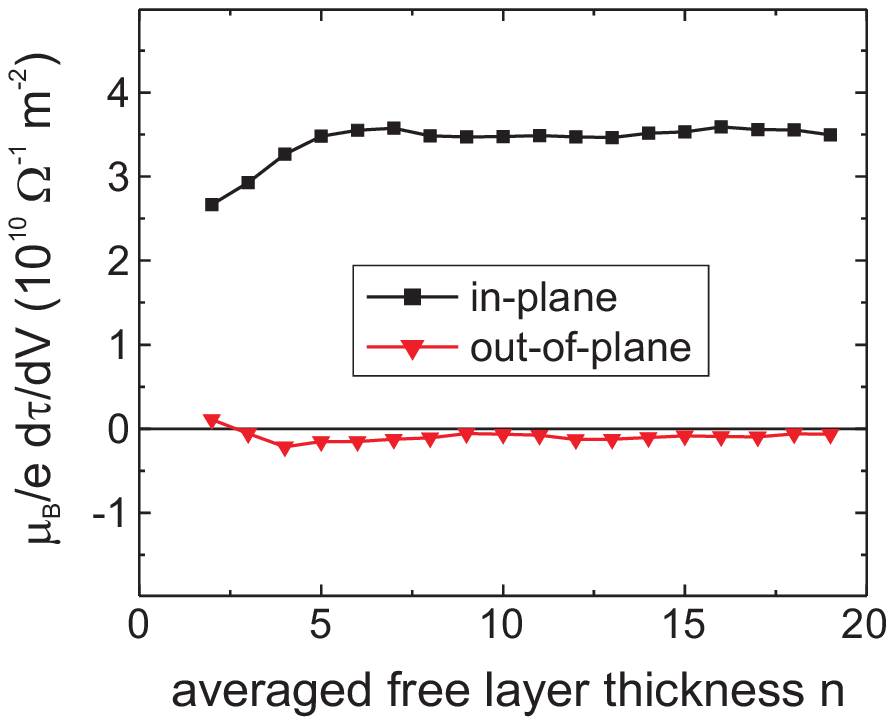}
\caption{
(Color online) Thickness fluctuations. Left: Sketch of assumed macroscopic
 fluctuation of Fe layer thickness. Fixed layer consists of equal parts 19
 monolayers and 20 monolayers of Fe. The free layer consists 
of $50\ \%\ n$, $25\ \%\ 
 (n+1)$, and $25\ \%\ (n-1)$ Fe monolayers. Right: Torkance as a function
 of the averaged free layer thickness $n$ assuming that each of the
 six possible junctions can be described separately. The total
 torkance is calculated by conducting the junctions in parallel. Note
 that the scale is identical to the scale in the bottom left panel of 
 Fig.~\ref{fig_torkance}.
}
\label{fig_fluc}
\end{figure}

Figure~\ref{fig_torkance} shows that it is possible to get a significant
out-of-plane component of the torkance in tunnel junctions that are
close to ideal.
However, thickness fluctuations in a real junctions lead
to an averaged torkance. To simulate such fluctuations we assume the
structure shown in the left panel of Fig.~\ref{fig_fluc}. The right
panel shows the
corresponding averaged torkance.  Comparing to
Fig.~\ref{fig_torkance} shows two consequences of the thickness
fluctuations. First, the oscillation of the in-plane component vanishes
and second, the out-of-plane component itself vanishes. 
This behavior is consistent with the experimental results in
Ref.~\onlinecite{sankey07}. Therefore, to actually measure a
non-vanishing out-of-plane component of the torkance or an oscillating
behavior of the in-plane component requires samples that are close 
to ideal. For
most samples the out-of-plane component will vanish.

In conclusion, we show that the spin-transfer torque in Fe/MgO/Fe
tunnel junctions behaves very similarly to the behavior found in
all-metallic devices.  The spin transfer torque is largely localized to
the interfaces and largely in the plane defined by the two
magnetizations.  The dominant contribution to the tunneling and the
torque comes from states around the Brillouin zone center where Fe is
a half-metal with respect to the $\Delta_1$ states. This half-metallic
behavior leads to an exponential decay of the torque within the
ferromagnetic layer even without dephasing.  
For a perfect sample we expect some small out-of-plane component and an
oscillation of the in-plane component of the torkance as a function of
the free layer thickness. However, small fluctuations of the thickness 
will average out this component.

\begin{acknowledgments}
This work has been supported in part by the NIST-CNST/UMD-NanoCenter
Cooperative Agreement.  We thank Dan Ralph and Paul Haney for 
useful conversations.
\end{acknowledgments}
%
%
%


\end{document}